\documentclass[epsfig]{article}
\usepackage{graphicx}
\usepackage{amsmath}

\input epsf.sty
\newtheorem{theorem}{Theorem}
\newtheorem{acknowledgement}[theorem]{Acknowledgement}

\input{tcilatex}

\begin{document}

\title{\rightline{\mbox {\normalsize {Lab/UFRHEP/0301-GNPHE/0301}}} \textbf{%
Geometric Engineering of }$\mathcal{N}\mathbf{=2}$\textbf{\ CFT}$_{4}$%
\textbf{s based on Indefinite Singularities: Hyperbolic Case}}
\author{M. Ait Ben Haddou$^{1,2,3}$\thanks{%
aitbenha@fsmek.ac.ma}, A. Belhaj$^{2,4}$\thanks{%
E-mail: adil.belhaj@uam.es} and E.H. Saidi$^{1,2}$\thanks{%
E-mail: H-saidi@fsr.ac.ma} \\
%EndAName
{\small 1 Lab/UFR-Physique des Hautes Energies, Facult\'{e} des Sciences de
Rabat, Morocco.}\\
{\small 2-Groupement National de Physique des Hautes Energies, GNPHE; Siege
focal, Rabat, Morocco.}\\
{\small 3} {\small D\'{e}partement de Math\'{e}matique \& Informatique,
Facult\'{e} des Sciences, Meknes, Morocco.}\\
{\small 4-Instituto de Fisica Teorica, C-XVI, Universidad Autonoma de
Madrid, E-28049-Madrid, Spain.}}
\maketitle

\begin{abstract}
Using Katz, Klemm and Vafa geometric engineering method of $\mathcal{N}=2$
supersymmetric QFT$_{4}$s and results on the classification of generalized
Cartan matrices of Kac-Moody (KM) algebras, we study the un-explored class
of $\mathcal{N}=2$ CFT$_{4}$s based on \textit{indefinite} singularities. We
show that the vanishing condition for the general expression of holomorphic
beta function of $\mathcal{N}=2$ quiver gauge QFT$_{4}$s coincides exactly
with the fundamental classification theorem of KM algebras. Explicit
solutions are derived for mirror geometries of CY threefolds with \textit{%
hyperbolic} singularities.
\end{abstract}

\tableofcontents

\bigskip

\bigskip

\begin{quote}
\textbf{Keywords}: \textit{Geometric engineering of }$\mathcal{N}\mathit{=2}$%
\textit{\ QFT}$_{4}$s\textit{, Indefinite and Hyperbolic Lie algebras, CY
threefolds with indefinite singularities, }$\mathcal{N}\mathit{=2}$\textit{\
CFT}$_{4}$s embedded in type II strings.

%\bigskip
\end{quote}

\thispagestyle{empty} \newpage \thispagestyle{empty} \newpage %
\setcounter{page}{1}

\newpage \newpage

\section{Introduction}

\qquad During the last few years supersymmetric $d$ dimension conformal
field theories ( CFT$_{d}$) have been subject to an intensive interest in
connection with superstring compactifications on Calabi-Yau manifolds ( CY)
\cite{1,2,3,4} and AdS/CFT correspondence \cite{5,6,7,8,9}. An important
class of these CFTs correspond to those embedded in type II string
compactifications on elliptic fibered CY threefolds with $ADE$ singularities
preserving eight supersymmetries. These field models, which give exact
solutions for the moduli space of the Coulomb branch and which admit a very
nice geometric engineering \cite{10} in terms of quiver diagrams, were shown
to be classified into two categories according to the type of
singularities:\ (i) $\mathcal{N}=2$ CFT$_{4}$, based on \textit{finite} $ADE$
singularities; with gauge group $G=\prod_{i}SU\left( n_{i}\right) $ and
matters in both fundamental $\mathbf{n}_{i}$ and bi-fundamental $\left(
\mathbf{n}_{i}\mathbf{,}\overline{\mathbf{n}}_{j}\right) $ representations
of $G$. (ii) $\mathcal{N}=2$ CFT$_{4}$ with gauge group $G=\prod_{i}SU\left(
s_{i}n\right) $ and bi-fundamental matters only. This second category of
scale invariant field models are classified by \textit{affine} $ADE$ Lie
algebras. The positive integers $s_{i}$ appearing in $G$ are the usual
Dynkin weights; they form a special positive definite integer vector $%
\mathbf{s}=\left( s_{i}\right) $ satisfying $\mathbf{K}_{ij}s_{j}=0$ and so
\begin{equation}
\mathbf{K}_{ij}n_{j}=0,  \label{1}
\end{equation}
where $n_{j}=ns_{j}=$ and where $\mathbf{K}$ is the Cartan matrix. The
appearance of this remarkable eq in the geometric engineering of $\mathcal{N}%
=2$ CFT$_{4}$s is very exciting; first because $4d$ conformal invariance,
requiring the vanishing of the holomorphic beta function, is now translated
into a condition on allowed Kac-Moody-Lie algebras eq(\ref{1}). Second, even
for $\mathcal{N}=2$ CFT$_{4}$ based on \textit{finite} $ADE$ with $m_{i}%
\mathbf{n}_{i}$ fundamental matters, the condition for scale invariance may
be also formulated in terms of $\mathbf{K}$\ as,
\begin{equation}
\mathbf{K}_{ij}n_{j}=m_{i}.  \label{2}
\end{equation}
The identities (\ref{1},\ref{2}) can be rigorously derived by starting from
mirror geometry of type IIA string on Calabi-Yau threefolds and taking the
field theory limit in the weak gauge couplings $g_{r}$ regime associated
with large volume base ( $V_{r}=1/\varepsilon $ with $\varepsilon
\rightarrow 0$ ). In this limit, one shows that complex deformations $%
a_{r,l} $ of the mirror geometry scale as $\varepsilon ^{l-k_{r}-1}$ and the
universal coupling parameters $Z^{\left( g_{r}\right) }$ ( given by special
ratios of $a_{r,l}$ and $a_{r\pm 1,l}$) behave as $\varepsilon ^{-b_{r}}$
where $b_{r}$ is the beta function coefficient of the $r$-th $U\left(
n_{r}\right) $\ gauge sub-group factor of $G$. Scale invariance of the CFT$%
_{4}$s requires $b_{r}=0$ $\forall r$\ and turns out to coincide exactly
with eqs(\ref{1},\ref{2}) depending on the type of $ADE$ singularities one
is considering; i.e affine or finite. The third exciting feature we want to
give here is that both eqs(\ref{1},\ref{2}) can be viewed as just the two
leading relations of the triplet
\begin{equation}
\mathbf{K}_{ij}^{\left( q\right) }n_{j}=qm_{i};\qquad q=0,+1,-1.  \label{3}
\end{equation}
The extra relation namely $\mathbf{K}_{ij}^{\left( -\right) }n_{j}=-m_{i}$
describes the so called \textit{indefinite} subset in the classification of
Kac-Moody-Lie algebras in terms of generalized Cartan matrices. If one
forgets for a while the algebraic geometry aspect of singular surfaces and
focus on the classification of $\mathcal{N}=2$ CFT$_{4}$s listed above, one
learns from the striking resemblance between the three sectors of eqs(\ref{3}%
) that it is legitimate to ask why not a third class of $\mathcal{N}=2$ CFT$%
_{4}$s associated with the third sector of eq(\ref{3}). A positive answer to
this question will not only complete the picture on $\mathcal{N}=2$ CFT$_{4}$%
s embedded in type II strings on CY threefolds given in \cite{10}; but will
also open an issue to approach singularities based on \textit{indefinite}
Lie algebras.

\qquad The triplet (\ref{3}) is not the unique motivation for our interest
in this third kind of $\mathcal{N}=2$\ supersymmetric scale invariant field
theory. There is also an other strong support coming from geometric
engineering of $\mathcal{N}=2$ QFT$_{4}$ \cite{10,11},\cite{12}. There, the
geometric engineering of fundamental matters requires the introduction of
the so called trivalent vertex. This vertex has a Mori vector $q_{\mathrm{%
\tau }}=\left( 1,-2,1;1,-1\right) $ with five entries ( sometimes called
also CY charges ). The first three ones are common as they are involved in
the geometric engineering of gauge fields and bi-fundamental matters. They
lead to eq(1). The fourth entry is used in the engineering of fundamental
matters of CFT$_{4}$s based on \textit{finite} $ADE$ and lead to eq(2). But
until now, the fifth entry has been treated as a spectator only needed to
ensure the CY condition $\sum_{\tau =1}^{5}q_{\mathrm{\tau }}=0$. Handling
this vertex on equal footing as the four previous others gives surprisingly
the missing third sector of eqs(\ref{3}) we are after.

\qquad In this paper, we study the remarkable class of the yet un-explored $%
\mathcal{N}=2$ CFT$_{4}$s associated with the third sector of eq(\ref{3}).
To achieve this goal, we have to prove that this kind of $\mathcal{N}=2$
CFTs really exist by computing the general expression of the holomorphic
beta function and then show that the moduli space of their solutions is non
trivial. To do so, we have to develop the following: (i) study \textit{%
indefinite} singularities of CY threefolds in relation with the third sector
of eqs(3). (ii) develop the geometric engineering of $\mathcal{N}=2$ QFT$%
_{4} $s embedded in type II strings on these class of singular CY 3-folds
and then solve $\mathcal{N}=2$ QFT$_{4}$ scale invariance constraint eqs.
Moreover, as the algebraic geometry of CY manifolds with \textit{indefinite}
singularities, like the classification of indefinite Lie algebras, are still
open questions, we will approach our problem throughout explicit examples.
Fortunately, we dispose actually of partial results dealing the
classification of a subset of indefinite Lie algebras. This concerns the so
called \textit{hyperbolic} subset having Dynkin diagrams very closed to the
usual \textit{finite} and \textit{affine }Lie algebras. Since we know much%
\textit{\ }about\textit{\ finite and affine }$ADE$ geometries, one suspects
to get here also exact results for hyperbolic mirror geometries. In addition
to the explicit results one expects, the study of hyperbolic singularities
is particularly interesting because it will give us more insight on the way
Katz, Mayr and Vafa basic result regarding $ADE$ bundles on elliptic curve
could be generalized to include indefinite singularities.

\qquad The presentation of this paper is as follows: In section $2$, we
review the computation of the general expression of beta function of $%
\mathcal{N}=2$ QFT$_{4}$s based on trivalent geometry. Here we show that the
general solutions for $\mathcal{N}=2$ CFT$_{4}$ scale invariance condition
coincides exactly with the Lie algebraic classification eq(\ref{3}). The
missing sector of $\mathcal{N}=2$ CFT$_{4}$s turns out to be intimately
related to the extra fifth entry of $q_{\mathrm{\tau }}$. In section $3$, we
study indefinite Lie algebras and too particularly their hyperbolic subset
as well links with $\mathcal{N}=2$ supersymmetric field theories in $4$
dimensions. In section $4$, we expose our explicit solutions for $\mathcal{N}%
=2$ CFT$_{4}$s based on hyperbolic singularities and in section $5$, we give
the conclusion and perspectives.

\section{Trivalent geometry and Beta function}

\qquad In \cite{10}, trivalent geometry has been introduced to engineer $%
\mathcal{N}=2$ supersymmetric QFT$_{4}$s ( with fundamental matter )
embedded in type IIA string theory on CY threefolds with $ADE$
singularities. This geometry extends the standard $ADE$ Dynkin diagrams and
involves higher dimension vertices \cite{10,12}. In this section, we give
the necessary tools one needs for the computation of the holomorphic beta
function of $\mathcal{N}=2$ QFT$_{4}$s. Details and techniques regarding
this geometry can be found in \cite{10,11}\ and subsequent works on this
subject \cite{12,13},.

\subsection{Trivalent geometry}

\qquad To illustrate the ideas, we start by considering the case of a unique
trivalent vertex; then we give the results for chains of trivalent vertices.

\textbf{a}) \textbf{Case of one trivalent vertex}

\qquad Since trivalent vertices depend on the kind of the Dynkin diagrams
one is using, we will fix our attention on those appearing in linear chains
ivolving $A_{k}$ type singularities. A quite similar analysis is also valid
for $DE$ graphs. In the case of $A$ type diagrams, trivalent geometry is
described by the typical three dimensional vertices $V_{i}$,
\begin{equation}
V_{0}=\left( 0,0,0\right) ;\quad V_{1}=\left( 1,0,0\right) ;\quad
V_{2}=\left( 0,1,0\right) ;\quad V_{3}=\left( 0,0,1\right) ;\quad
V_{4}=\left( 1,1,1\right)  \label{4}
\end{equation}
satisfying the following toric geometry relation
\begin{equation}
\sum_{i=0}^{4}q_{i}V_{i}=-2V_{0}+V_{1}+V_{2}+V_{3}-V_{4}=0  \label{5}
\end{equation}
The vector charge $\left( q_{i}\right) =\left( -2,1,1,1,-1\right) $ is known
as the Mori vector and the sum of its $q_{i}$ components is zero as required
by the CY condition; $\sum_{i}q_{i}=0$. In type IIB strings on mirror CY3,
the $\left( V_{0},V_{1},V_{2},V_{3},V_{4}\right) $ vertices are\ represented
by complex variables $\left( u_{0},u_{1},u_{2},u_{3},u_{4}\right) $
constrained as $\prod_{i}u_{i}^{q_{i}}=1$ and solved by $\left(
1,x,y,z,xyz\right) $; see figure 1. \bigskip
\begin{figure}[tbh]
\begin{center}
\epsfxsize=4cm \epsffile{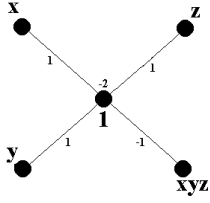}
\end{center}
\caption{{\protect\small \textit{This figure represents a typical vertex in
trivalent mirror geometry. To the central node, it is attached four legs;
two of them are of Dynkin type and the two others are linked to CY toric
geometry and deals, amongst others, with fundamental matters.}}}
\end{figure}
In terms of these variables, the algebraic eq describing mirror geometry,
associated to eq(\ref{5}), is given by the following complex surface,
\begin{equation}
P\left( X^{\ast }\right) =\mathrm{e}+\mathrm{a}x+\mathrm{b}y+\left( \mathrm{c%
}-\mathrm{d}xy\right) z,  \label{6}
\end{equation}
where $\mathrm{a,b,c,d}$ and $\mathrm{e}$ are complex moduli. Note that upon
eliminating the $z$ variable, the above ( trivalent) algebraic geometry eqs
reduces exactly to the standard bivalent vertex of $A_{1}$ geometry,
\begin{equation}
P\left( X^{\ast }\right) =\mathrm{a}x+\mathrm{e}+\frac{\mathrm{bc}}{\mathrm{d%
}}\frac{1}{x}.  \label{7}
\end{equation}
The monomials $y_{0}=x$, $y_{1}=1$ and $y_{2}=\frac{1}{x}$ satisfy the well
known $su\left( 2\right) $ relation namely $y_{0}y_{2}=y_{1}^{2}$. To get
algebraic geometry eq of the CY3, one promotes the non zero coefficients $%
\mathrm{a,b,c,d}$ and $\mathrm{e}$ to holomorphic polynomials a $\mathbf{CP}%
^{1}$ as follows:

\begin{eqnarray}
\mathrm{e} &=&\sum_{i=0}^{n_{r}}\mathrm{e}_{i}w^{i};\qquad \mathrm{a}%
=\sum_{i=0}^{n_{r-1}}\mathrm{a}_{i}w^{i};\qquad \mathrm{b}%
=\sum_{i=0}^{n_{r+1}}b_{i}w^{i},  \notag \\
\mathrm{c} &=&\sum_{i=0}^{m_{r}}c_{i}w^{i};\qquad \mathrm{d}%
=\sum_{i=0}^{m_{r}^{\prime }}d_{i}w^{i};\qquad \mathrm{e}_{0},\mathrm{a}_{0},%
\mathrm{b}_{0},\mathrm{c}_{0},\mathrm{d}_{0}\neq 0.  \label{8}
\end{eqnarray}
These analytic polynomials encode the fibrations of $SU\left(
1+n_{r-1}\right) \times SU\left( 1+n_{r}\right) \times SU\left(
1+n_{r+1}\right) $ gauge and $SU\left( 1+m_{r}\right) \times SU\left(
1+m_{r}^{\prime }\right) $ flavor symmetries; of the underlying $\mathcal{N}%
=2$ QFT$_{4}$; engineered over the nodes of the trivalent vertex. For
instance $SU\left( 1+n_{r-1}\right) $ gauge symmetry is fibered over $V_{0}$
and $SU\left( 1+m_{r}\right) $ and $SU\left( 1+m_{r}^{\prime }\right) $\
flavor invariances are fibered over the nodes $V_{3}$ and $V_{4}$
respectively, see figure2.

\begin{figure}[tbh]
\begin{center}
\epsfxsize=8cm \epsffile{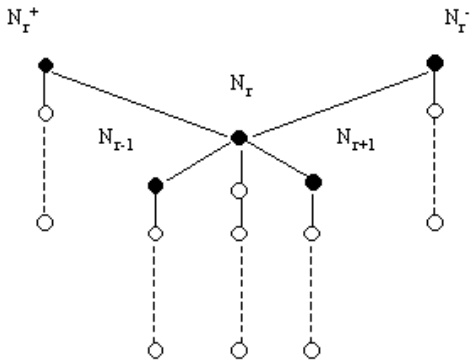}
\end{center}
\caption{{\protect\small \textit{This graph describes a typical vertex one
has in trivalent geometric engineering of }}$\mathcal{N}\mathit{=2}$
{\protect\small \textit{supersymmetric QFT}}$_{4}$. $SU\left( 1+l\right) $%
{\protect\small \textit{\ gauge and flavour symmetries are fibered over the
five black nodes. Flavor symmetries require large base volume.}}}
\end{figure}
Note that the functions $\mathrm{a,b,c,d}$ and $\mathrm{e}$ are not all of
them independent as one can usually fix one of them. In \cite{10}, the
coefficient $\mathrm{d}$ of eq(\ref{7}) was set to one and $\mathrm{c}$ kept
arbitrary in order to geometric engineer the needed fundamental matters for
\textit{finite} ADE CFT$_{4}$s. Here we will keep all of these moduli
arbitrary and fix one of them only at appropriate time. The reason is that
by fixing one of these moduli from the begining; one rules out a full sector
in the moduli space of CFT$_{4}$s as it has been the case for the CFT$_{4}$
sector associated with indefinite Lie algebras we want to study here.

$\mathcal{N}=2$ \textbf{QFT}$_{4}$\textbf{\ limit}

\qquad To get the various $\mathcal{N}=2$ CFT$_{4}$s embedded in type II
strings on CY3-folds, we have to study the field theory limit one gets from
mirror geometry of type IIA string on CY3 and look for the scaling
properties of the gauge coupling constants moduli. We will do this for the
case of one trivalent vertex eq(\ref{6}) and then give the general result
for a chain of several trivalent vertices. To that purpose, we proceed in
three steps: First determine the behaviour of the complex moduli $\mathrm{f}%
_{i}$ appearing in the expansion eq(\ref{8}) under a shift of $w$ by $%
1/\varepsilon $ with $\varepsilon \rightarrow 0$. Doing this and requiring
that eqs(\ref{8}) should be preserved, that is still staying in the
singularity described by eqs(\ref{8}), we get the following,

\begin{eqnarray}
\mathrm{e}_{l} &\sim &\varepsilon ^{l-n_{r}};\qquad \mathrm{a}_{l}\sim
\varepsilon ^{l-n_{r-1}};\qquad \mathrm{b}_{l}\sim \varepsilon ^{l-n_{r+1}};
\notag \\
\mathrm{c}_{l} &\sim &\varepsilon ^{l-m_{r}};\qquad \mathrm{d}_{l}\sim
\varepsilon ^{l-m_{r}^{\prime }}.  \label{9}
\end{eqnarray}
Second compute the scaling of the\ gauge coupling constant moduli $Z^{\left(
g\right) }$ under the shift $w^{\prime }=w+1/\varepsilon $. Putting eqs(\ref
{9}) back into the explicit expression of $Z^{\left( g\right) }$ namely
\begin{equation}
Z^{\left( g\right) }=\frac{a_{0}b_{0}c_{0}}{e_{0}^{2}d_{0}},  \label{10}
\end{equation}
we get the following behaviour $Z^{\left( g\right) }\sim \varepsilon
^{-b_{r}}$ with $b_{r}$ given by,
\begin{equation}
b_{r}=2n_{r}-n_{r-1}-n_{r+1}-\left( m_{r}-m_{r}^{\prime }\right) .
\label{11}
\end{equation}
In the limit $\varepsilon \rightarrow 0$, finiteness of $Z^{\left( g\right)
} $ requires then that the field theory limit should be asymptotically free;
that is $b_{r}\leq 0$. Satured bound $b_{r}=0$ corresponds to scale
invariance we are interested in here. With these relations at hand, let us
see how they extend to the case of several trivalent vertices.

\textbf{b}) \textbf{Chains of trivalent vertices}

\qquad Thinking about the vertices of eq(\ref{5}) as a generic trivalent
vertex of a linear chain of $N$ trivalent vertices, that is
\begin{eqnarray}
V_{0} &\rightarrow &V_{\alpha }^{0};\qquad V_{3}\rightarrow V_{\alpha
}^{+};\qquad V_{4}\rightarrow V_{\alpha }^{-}  \notag \\
V_{1} &\rightarrow &V_{\alpha -1}^{0};\qquad V_{2}^{0}\rightarrow V_{\alpha
+1}^{0},  \label{12}
\end{eqnarray}
and varying $\alpha $\ on the set $\left\{ 1,...,N\right\} $\ together with
intersections between $V_{\alpha }^{0}$ and $V_{\alpha \pm 1}^{0}$ specified
by Mori vector $q_{\alpha }^{i}$, one can build more general toric
geometries. In the generic case, the analogue of eq(\ref{5}) extends as
\begin{equation}
\sum_{\alpha \geq 0}\left( q_{\alpha }^{i}V_{\alpha
}^{0}+V_{i}^{+}-V_{i}^{-}\right) =0.  \label{13}
\end{equation}
Note that the $\pm $ upper indices carried by the $V_{i}^{\pm }$ vertices
refer to the fourth $+1$ and five $-1$ entries of the Mori vector $q_{\tau
}^{i}=\left( q_{\alpha }^{i};+1,-1\right) $ of trivalent vertex. In
practice, the Mori vectors $q_{\alpha }^{i}$s form a $N\times \left(
N+s\right) $ rectangular matrix whose $N\times N$ square sub-matrix $%
q_{j}^{i}$ is minus the generalized Cartan matrix $K_{ij}$. For the example
of affine $A_{N-1}$, the Mori charges read as $q_{\alpha }^{i}=2\delta
_{\alpha }^{i}-\delta _{\alpha }^{i-1}-\delta _{\alpha }^{i+1}$ with the
usual periodicity of affine $SU\left( N\right) $. The remaining $N\times s$
part of $q_{\alpha }^{i}$ is fixed by the Calabi-Yau condition $\sum_{\alpha
}q_{\alpha }^{i}=0$ and the corresponding vertices are interpreted as
dealing with non compact divisors defining the singular space on which live
singularities. In mirror geometry where $x_{\alpha -1},$\ $x_{\alpha }$,$\
x_{\alpha +1},\ y_{\alpha },$\ and $\frac{x_{\alpha -1}x_{\alpha
+1}y_{\alpha }}{y_{\alpha }^{2}}$\ are the variables associated with the
vertices (\ref{12}), eq(\ref{6}) extends as \textrm{a}$_{\alpha -1}x_{\alpha
-1}+\mathrm{a}_{\alpha }x_{\alpha }$+\textrm{a}$_{\alpha +1}x_{\alpha +1}+%
\mathrm{c}_{\alpha }y_{\alpha }+\mathrm{d}_{\alpha }\frac{x_{\alpha
-1}x_{\alpha +1}y_{\alpha }}{x_{\alpha }^{2}}=0$ where $\mathrm{a}_{\alpha }$%
, $\mathrm{c}_{\alpha }$ and $\mathrm{d}_{\alpha }$ are complex moduli.
Summing over the vertices of the chain and setting $y_{\alpha }=x_{\alpha
}z_{\alpha }$, one gets
\begin{equation}
P\left( X^{\ast }\right) =\mathrm{a}_{0}x_{0}+\sum_{\alpha \geq 1}\left(
\mathrm{a}_{\alpha }x_{\alpha }+\mathrm{c}_{\alpha }x_{\alpha }z_{\alpha }+%
\mathrm{d}_{\alpha }\frac{x_{\alpha -1}x_{\alpha +1}z_{\alpha }}{x_{\alpha }}%
\right) .  \label{130}
\end{equation}
Eliminating the variable $z_{\alpha }$ as we have done for eq(\ref{7}), we
obtain
\begin{equation}
P\left( X^{\ast }\right) =\sum_{\alpha \geq 0}x^{\alpha }\mathrm{a}_{\alpha
}\left( w\right) \prod_{\beta \geq 1}\left( \frac{\mathrm{c}_{\beta }\left(
w\right) }{\mathrm{d}_{\beta }\left( w\right) }\right) ^{\alpha -\beta }.
\label{131}
\end{equation}
We will use this expression as well as eqs(\ref{8}) when building mirror
geometries of CY threefolds with hyperbolic singularities.

\subsection{Classification of $\mathcal{N}=2$ CFT$_{4}$s}

\qquad Using eq(\ref{11}), and focusing on the interesting situation where
\begin{equation}
n_{\alpha }=\text{\ }m_{\alpha }=m_{\alpha }^{\prime }=0\text{ \qquad for \
\ }\alpha >N,  \label{14}
\end{equation}
and all remaining others are non zero, the condition for conformal
invariance of the underlying $\mathcal{N}=2$ supersymmetric gauge theory
reads, in terms of generalized Cartan matrices $\mathbf{K}$ of Lie algebras,
as
\begin{equation}
\mathbf{K}_{ij}n_{j}-\left( m_{i}-m_{i}^{\prime }\right) =0;\qquad 1\leq
i\leq N.  \label{15}
\end{equation}
This is a very remarkable relation; first because it can be put into the
form eq(\ref{3}) and second its solutions, which depend on the sign of $%
\left( m_{i}-m_{i}^{\prime }\right) $, are exactly given by the fundamental
theorem on the classification of Lie algebras. Let us first recall this
theorem and then give our solutions

\textbf{a}) \textbf{Theorem I: Classification of Lie algebras}\newline
\qquad A generalized indecomposable Cartan matrix $\mathbf{K}$ obey one and
only one of the following three statements:\newline
\textbf{i}) \textit{Finite type ( }$\det \mathbf{K}>0$ ): There exist a real
positive definite vector $\mathbf{u}$ ( $u_{i}>0;$ $i=1,2,...$) such that
\begin{equation}
\mathbf{K}_{ij}u_{j}=v_{j}>0.  \label{16}
\end{equation}
\textbf{ii}) \textit{Affine type, }corank$\left( \mathbf{K}\right) =1$, $%
\det \mathbf{K}=0$\textit{: }There exist a unique, up to a multiplicative
factor, positive integer definite vector $\mathbf{n}$ ( $n_{i}>0;$ $%
i=1,2,... $) such that
\begin{equation}
\mathbf{K}_{ij}n_{j}=0.  \label{17}
\end{equation}
\textbf{iii}) \textit{Indefinite type ( }$\det \mathbf{K}\leq 0$ ), corank$%
\left( \mathbf{K}\right) \neq 1$\textit{: }There exist a real positive
definite vector $\mathbf{u}$ ($u_{i}>0;$ $i=1,2,...$) such that
\begin{equation}
\mathbf{K}_{ij}u_{j}=-v_{i}<0.  \label{18}
\end{equation}
Eqs(\ref{16}-\ref{18}) combine together to give eq(\ref{3}). These eqs tell
us, amongst others, that whenever there exist a real ( integer ) positive
definite vector $u_{i}>0$ such that $\mathbf{K}_{ij}u_{j}<0$, then the
Cartan matrix is of indefinite type.

\textbf{b}) \textbf{Solutions of eq}(\ref{15})\newline
\qquad Setting $u_{i}=n_{i}$ and $v_{i}=\left| m_{i}-m_{i}^{\prime }\right| $
in the constraint eq(\ref{15}) required by the vanishing of the beta
function, one gets the general solutions for $\mathcal{N}=2$ supersymmetric
conformal invariance in four dimensions. The previous theorem teaches us
that there should exist \textit{three} kinds of $\mathcal{N}=2$ CFT$_{4}$s
in one to one correspondence with \textit{finite}, \textit{affine} and
\textit{indefinite} Lie algebras. These CFTs are as follows:

(\textbf{i}) $\mathcal{N}=2$ CFT$_{4}$s based on \textit{finite} Lie
algebras; this subset is associated with the case $m_{i}>m_{i}^{\prime }\geq
0$.

(\textbf{ii}) $\mathcal{N}=2$ CFT$_{4}$s based on \textit{affine} Lie
algebras; they correspond to the case $m_{i}=m_{i}^{\prime }$ and too
particularly $m_{i}=m_{i}^{\prime }=0$ considered in QFT$_{4}$ literature.

(\textbf{iii}) $\mathcal{N}=2$ CFT$_{4}$s based on \textit{indefinite} Lie
algebras; they correspond to the case $0\leq m_{i}<m_{i}^{\prime }$. \newline
This result generalizes the known classification concerning $\mathcal{N}=2$
CFT$_{4}$s using \textit{finite} and \textit{affine} Lie algebras which are
recovered here by setting $m_{i}^{\prime }=0$. For the other remarkable case
where $m_{i}=0$; but $m_{i}^{\prime }>0$, we get the missing third class of $%
\mathcal{N}=2$ CFT$_{4}$s involving indefinite Lie algebras. Before giving
the geometric engineering of these $\mathcal{N}=2$ CFT$_{4}$s, note that the
beta function eqs(\ref{11}) is invariant under the change $m_{i}\rightarrow
m_{i}-m_{i}^{\ast }$ and $m_{i}^{\prime }\rightarrow m_{i}^{\prime
}-m_{i}^{\ast }$. By appropriate choices of the positive numbers $%
m_{i}^{\ast }$; for instance by taking $m_{i}^{\ast }=m_{i}$\ or setting $%
m_{i}^{\ast }=m_{i}^{^{\prime }}$ one can usually rewrite eq(\ref{15}) as in
eq(\ref{3}) namely $\mathbf{K}_{ij}^{\left( q\right) }n_{j}=qm_{i}$; with $%
q=0,\pm 1$. This symmetry reflects just the freedom to fix one of the
complex moduli of eq(\ref{130}).

\section{More on Indefinite Lie algebras}

\qquad As the subject of indefinite Lie algebras is still a mathematical
open problem since the full classification of their generalized Cartan
matrices has not yet been achieved, we will focus our attention here on the
Wanglai Li special subset \cite{14,15},\cite{16}, known also as \textit{%
hyperbolic} Lie algebras. This is a subset of indefinite Kac-Moody-Lie
algebras which is intimately related to finite and affine ones and on which
we know much about their classification. The results we will derive here
concerns these hyperbolic algebras; but they apply as well to other kinds of
indefinite Lie algebras that are not of Wanglai Li type. For other
applications of hyperbolic Lie algebras in string theory; see \cite{17,18} -
\cite{19,20}.

\qquad To start, note that the derivation of \textit{hyperbolic} Lie
algebras is based on the same philosophy one uses in building affine Lie
algebras $\widehat{\mathrm{g}}$ from finite ones $\mathrm{g}$ by adding a
node to the Dynkin diagram of \textrm{g}. Using this method, Wanglai Li
constructed and classified the $238$\ possible Dynkin diagrams of the
hyperbolic Lie algebras from which one derives their generalized Cartan
matrices. The corresponding diagrams, which were denoted in \cite{14} as $%
\mathcal{H}_{i}^{n}$; $i=1,...$, contain as a sub-diagram of co-order $1$,
the usual Dynkin graphs associated with $\widehat{\mathrm{g}}$ and $\mathrm{g%
}$ Lie algebras. In other words, by cutting a node of an order $n$ \textit{%
hyperbolic} Dynkin diagram, the resulting $\left( n-1\right) $-\textit{th}\
sub-diagrams one gets is one of the two following: (i) either it coincides
with one of the Dynkin graphs of $\mathrm{g}$; or (ii) it coincides with an
affine $\widehat{\mathrm{g}}$ one. The general structure of \textit{%
hyperbolic Dynkin diagrams} are then of the form figure 3;

\bigskip

\bigskip

\begin{figure}[tbh]
\begin{center}
\epsfxsize=8cm \epsffile{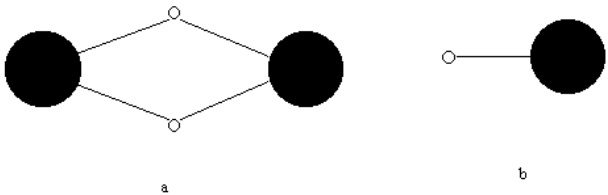}
\end{center}
\caption{{\protect\small \textit{Figures 3a and 3b represent the two kinds
of Dynkin diagrams of Hyperbolic Lie algebras. These diagrams are built by
adding one node to given finite and affine Dynkin graphs. The full set of
hyperbolic Dynkin diagrams may be found in \protect\cite{18}.}}}
\end{figure}

In what follows we give some features of $\mathcal{H}_{i}^{n}$ hyperbolic
algebras that are close to finite and affine $ADE$.

\subsection{Hyperbolic Lie algebras}

\qquad Following \cite{14,15}-\cite{16}, an indecomposable generalized
Cartan matrix $\mathbf{K}$ is said to be \textit{hyperbolic} (resp \textit{%
strictly hyperbolic}) if: (1) it is of \textit{indefinite} type and (2) any
connected proper sub-diagram of the Dynkin diagram, associated with the
matrix $\mathbf{K}$, is of finite or affine ( rep \textit{finite} ) type.

\textbf{Classification}\newline
\qquad Motivated by physical applications, we prefer to rearrange the $%
\mathcal{H}_{i}^{n}$ Wanglai Li hyperbolic algebras into two types as
follows:\newline
(\textbf{1}) \textit{Type I hyperbolic Lie algebras,} ($TypeI\mathcal{H}$).
They concern the $\mathcal{H}_{i}^{n}$s that contain order $\left(
n-1\right) $\ \textit{affine }Lie algebras. These $\mathcal{H}_{i}^{n}$
algebras are given by the following list:\newline

\begin{itemize}
\item  \textit{Simply laced} $TypeI\mathcal{H}$: A careful inspection of the
Wanglai Li classification shows that simply laced diagrams are:
\begin{eqnarray}
&&\mathcal{H}_{1}^{4},\quad \mathcal{H}_{2}^{4},\quad \mathcal{H}%
_{3}^{4},\quad \mathcal{H}_{1}^{5},\quad \mathcal{H}_{8}^{5},\quad \mathcal{H%
}_{1}^{6},\quad \mathcal{H}_{5}^{6},\quad \mathcal{H}_{6}^{6},\quad \mathcal{%
H}_{1}^{7},  \notag \\
&&\mathcal{H}_{4}^{7},\quad \mathcal{H}_{1}^{8},\quad \mathcal{H}%
_{4}^{8},\quad \mathcal{H}_{5}^{8},\quad \mathcal{H}_{1}^{9},\quad \mathcal{H%
}_{4}^{9},\quad \mathcal{H}_{5}^{9},\quad \mathcal{H}_{1}^{10},\quad
\mathcal{H}_{4}^{10}.  \label{190}
\end{eqnarray}
These hyperbolic algebras contain the well known affine $ADE$ as maximal Lie
subalgebras; some of them have internal discrete automorphisms. Note in
passing that $\mathcal{H}_{i}^{n}$ Dynkin graphs with $D$ and $E$
sub-diagrams are denoted, in trivalent geometry as T$_{\left( p,q,r\right) }$
or again as $DE_{s}$.

\item  Non \textit{simply laced} $TypeI\mathcal{H}$: There are $63$ non
simply laced diagrams of $TypeI\mathcal{H}$; those having an order greater
than $6$ are as follows
\begin{equation}
\mathcal{H}_{2}^{7},\quad \mathcal{H}_{3}^{7},\quad \mathcal{H}%
_{2}^{8},\quad \mathcal{H}_{3}^{8},\quad \mathcal{H}_{2}^{9},\quad \mathcal{H%
}_{3}^{9},\quad \mathcal{H}_{2}^{10},\quad \mathcal{H}_{3}^{10},
\end{equation}
the others can be found in \cite{15}.
\end{itemize}

(\textbf{2}) \textit{Type II hyperbolic Lie algebras}, ($TypeII\mathcal{H}$%
). They concern the $\mathcal{H}_{i}^{n}$s that are not in $TypeI\mathcal{H}$
list; their Dynkin diagrams do contain no order $\left( n-1\right) $ affine
sub-diagram once cutting a node.

\textbf{Hyperbolic Symmetries}\newline
\qquad In studying Wanglai Li hyperbolic symmetries, one should distinguish
two cases: hyperbolic algebras of order two and those with orders greater
than two. For the first ones there are infinitely many, whereas the number
of the second type is finite. The two following theorems classify these
algebras.

\textbf{a}) \textsl{Theorem II: }Let $\mathbf{K}$ be an indecomposable
generalized Cartan matrix of order $2$ with $K_{11}=K_{22}=2$ and $K_{12}=-a$%
\ and $K_{21}=-b$ where $a$ and $b$ are positive integers; then, we have the
following classification

(i) $\mathbf{K}$ is of\textit{\ finite} type if and only if $det\mathbf{K>}0$%
; i.e $ab<4$

(ii) $\mathbf{K}$ is of \textit{affine} type if and only if $det\mathbf{K=}0$%
; i.e $ab=4$

(iii) $\mathbf{K}$ is of\textit{\ indefinite} type if and only if $det%
\mathbf{K<}0$; i.e $ab>4$\newline
From this result, one sees that while $ab\leq 4$ has a finite number of
solutions for positive integers $a$ and $b$, there are however infinitely
many for $ab>4$; but no one of the corresponding algebra is simply laced.
For orders greater than two, there is a similar classification; but the
following theorem shows that, for $n\geq 3$, there exist however a finite
set of hyperbolic algebras.

\textbf{b}) \textsl{Theorem III}\newline
\qquad The full list of the Dynkin diagrams of hyperbolic Cartan matrices
for $3\leq n\leq 10$ is given in \cite{14,15}; see also \cite{16}.There are
altogether $238$ hyperbolic diagrams, $35$ diagrams are strictly hyperbolic
ones and $142$ diagrams are symmetric or symmetrisable.

(i) The orders $n$ of \textit{strictly hyperbolic} Cartan matrices are
bounded as $2\leq n$ $\leq 5$

(ii) The orders $n$ of a \textit{hyperbolic} Cartan matrices are bounded as $%
2\leq n\leq 10$

\subsection{From Affine to Hyperbolic}

\qquad In this subsection, we study simply laced $typeI\mathcal{H}$
hyperbolic Lie algebras (\ref{190}); see also figure 2. This subset is more
a less simple to handle when building mirror geometries of $\mathcal{N}=2$
CFT$_{4}$s embedded in type IIA strings on CY Threefolds. Simply laced
hyperbolic geometries has no branch cuts; but can be also extended to the
case of general geometries associated with non simply laced hyperbolic
symmetries; especially those obtained from simply laced $typeI\mathcal{H}$
algebras by following the idea of folding used in \cite{12,21}-\cite{22}.
Moreover, as simply laced $typeI\mathcal{H}$ algebras contain affine $ADE$
symmetries as maximal subalgebras, we will use results on affine geometries
to derive the mirror geometries associated with $typeI\mathcal{H}$. To do
so, we start from the known results on affine $ADE$ geometries and look for
generalizations that solve the constraint eqs required by $typeI\mathcal{H}$
invariance. For pedagogical reasons, we will illustrate our method on the
four examples of figure 4.

\begin{figure}[tbh]
\begin{center}
\epsfxsize=14cm \epsffile{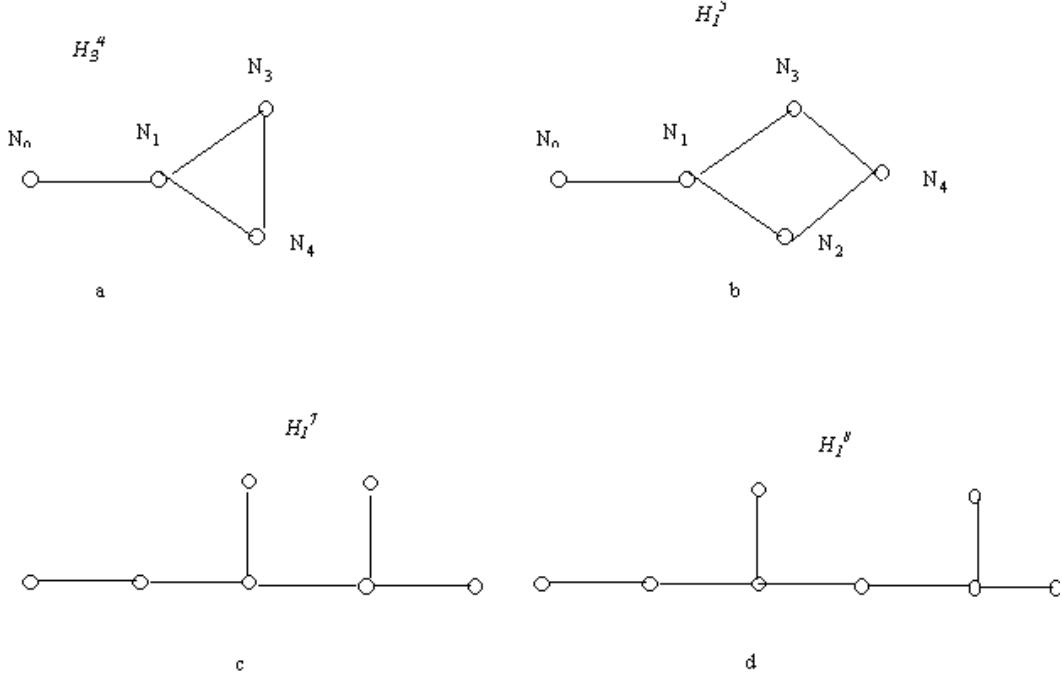}
\end{center}
\caption{{\protect\small \textit{Figures 4a, 4b, 4c and 4d represent four
examples of simply laced Dynkin diagrams of order three and four Hyperbolic
Lie algebras. The simple root system as well as Cartan matrices associated
with these graphs may be easily read from the }}$\widehat{\mathit{A}}_{2}$,
{\protect\small \textit{\ }}$\widehat{\mathit{A}}_{3}${\protect\small
\textit{\ and }}$\widehat{\mathit{D}}_{5}${\protect\small \textit{\ affine
ones.}}}
\end{figure}
Generalization to other diagrams is straightforward. Let us first give some
particular features on these special hyperbolic Lie algebras, then consider
the building of the corresponding mirror geometries and hyperbolic $\mathcal{%
N}=2$\ CFT$_{4}$s.

\begin{itemize}
\item  $\mathcal{H}_{3}^{4}$ \textit{hyperbolic Lie algebra}
\end{itemize}

From the Dynkin diagram of figure 4a, one derives the following generalized
Cartan matrix with diagonal entries $2$ associated with the $\left(
1+3\right) $ nodes $N_{i}$ and off diagonal negative integers describing
intersections
\begin{equation}
\mathbf{K}\left( \mathcal{H}_{3}^{4}\right) =\left(
\begin{array}{cccc}
2 & -1 & 0 & 0 \\
-1 & 2 & -1 & -1 \\
0 & -1 & 2 & -1 \\
0 & -1 & -1 & 2
\end{array}
\right) .  \label{21}
\end{equation}
This matrix $\mathbf{K}_{ij}$, $0\leq i,j\leq 3$ has a negative determinant (%
$det\mathbf{K}=-3$ ) and exhibits remarkable properties in agreement with
the classification of \cite{16}. $\mathbf{K}\left( \mathcal{H}%
_{3}^{4}\right) $ contains the Cartan matrix of affine $\mathbf{A}_{2}$\ and
that of finite $\mathbf{A}_{3}$ as $3\times 3$ sub-matrices.\ The first one
is obtained by subtracting the first row and column of (\ref{21}) or
equivalently by cutting the node $N_{0}$ of figure 4a. The second is
recovered by subtracting the last row and column which correspond to cutting
$N_{3}$.

\begin{itemize}
\item  $\mathcal{H}_{1}^{5}$ \textit{hyperbolic Lie algebra}

Figure 4b leads to the the following $\mathcal{H}_{1}^{5}$ generalized
Cartan matrix ,
\begin{equation}
\mathbf{K}\left( \mathcal{H}_{1}^{5}\right) =\left(
\begin{array}{ccccc}
2 & -1 & 0 & 0 & 0 \\
-1 & 2 & -1 & 0 & -1 \\
0 & -1 & 2 & -1 & 0 \\
0 & 0 & -1 & 2 & -1 \\
0 & -1 & 0 & -1 & 2
\end{array}
\right) .
\end{equation}
The determinant of this matrix is equal to $-4$ and its Dynkin diagram
contains those of affine $\mathbf{A}_{3}$, finite $\mathbf{A}_{4}$ and
finite $\mathbf{D}_{4}$ as three sub-graphs of order $4$. Like for $TypeI%
\mathcal{H}_{3}^{4}$, the Dynkin graph of $\mathcal{H}_{1}^{5}$ has a $Z_{2}$
internal automorphism fixing three nodes and interchanging two.

\item  $\mathcal{H}_{1}^{7}$ and $\mathcal{H}_{1}^{8}$\ \textit{hyperbolic
Lie algebras}
\end{itemize}

Denoted also as $DE_{7}$ in \cite{16}, the generalized Cartan matrix for $%
\mathcal{H}_{1}^{7}$ reads as,
\begin{equation}
\mathbf{K}\left( \mathcal{H}_{1}^{7}\right) =\left(
\begin{array}{cccccc}
2 & -1 & 0 & 0 & 0 & 0 \\
-1 & 2 & 0 & -1 & 0 & 0 \\
0 & 0 & 2 & -1 & 0 & 0 \\
0 & -1 & -1 & 2 & -1 & -1 \\
0 & 0 & 0 & -1 & 2 & 0 \\
0 & 0 & 0 & -1 & 0 & 2
\end{array}
\right) .
\end{equation}
This is a hyperbolic Lie algebra of order $7$ and has three order six
subalgebras namely affine $D_{5}$, finite $D_{6}$ and finite $E_{6}$.
Similarly, one can check that the order seven subalgebras of $\mathcal{H}%
_{1}^{8}$ ($\equiv DE_{8}$) are affine $D_{6}$, finite $D_{7}$ and finite $%
E_{7}$.

\section{Hyperbolic CFT$_{4}$s}

\qquad Following \cite{10}, mirror geometries of type IIA string on CY
threefolds with $ADE$ singularities are described by an algebraic geometry
eq of the form
\begin{equation}
P\left( X^{\ast }\right) =\sum_{\alpha }a_{\alpha }y_{\alpha },  \label{210}
\end{equation}
where $a_{\alpha }=a_{\alpha }\left( w\right) $ are complex moduli with
expansion of type eq(\ref{8}) and where the $y_{\alpha }$ complex variables
are constrained as
\begin{equation}
\prod_{j=1}^{n}y_{j}^{q_{j}^{i}}=\prod_{\alpha =n+1}^{n+4}y_{\alpha
}^{-q_{\alpha }^{i}}.  \label{211}
\end{equation}
In this eq $q_{j}^{i}$ is minus $\mathbf{K}_{ij}$ and $y_{\alpha }$, with $%
n<\alpha <n+5$, are some extra complex variables that define the elliptic
curve on which lives the singularity.

\subsection{Flat ADE bundles on elliptic curve}

\qquad In the case of affine $ADE$ singularities, the $n$ first variables $%
y_{i}$ are solved in terms of the the four extra ones, $y_{n+1}$, $y_{n+2}$,
$y_{n+3}$ and $y_{n+4}$, which themselves are realized as
\begin{equation}
y_{n+1}=y^{2},\quad y_{n+2}=x^{3},\quad y_{n+3}=z^{6},\quad y_{n+4}=xyz,
\label{2110}
\end{equation}
where $\left( y,x,z\right) $\ are the homogeneous coordinates of the
weighted projective space $\mathbf{WP}^{2}\left( 3,2,1\right) $.\ These
monomials generate an elliptic curve,
\begin{equation}
y^{2}+x^{3}+z^{6}+xyz=0,  \label{2111}
\end{equation}
embedded in $\mathbf{WP}^{2}\left( 3,2,1\right) \subset \mathbf{WP}%
^{3}\left( 3,2,1,6-N\right) $. Using $y,x,z$ and $v$ coordinates of $\mathbf{%
WP}^{3}\left( 3,2,1,6-N\right) $ and considering the generalized Cartan
matrix $\mathbf{K}_{ij}$ of an affine Lie algebra, one also gets by help of
eqs(\ref{211}) the other $y_{i}$s. Putting altogether in eqs(\ref{210}), we
obtain the following result for the cases of affine $\widehat{A}_{2}$ and
affine $\widehat{A}_{3}$ geometries respectively
\begin{eqnarray}
\widehat{A}_{2} &:&y^{2}+x^{3}+z^{6}+xyz+v\left( \mathrm{b}z^{3}+\mathrm{c}%
xz+\mathrm{d}y\right) =0,  \label{2112a} \\
\widehat{A}_{3} &:&y^{2}+x^{3}+z^{6}+xyz+v\left( \mathrm{b}z^{4}+\mathrm{c}%
xz^{2}+\mathrm{d}yz+\mathrm{e}x^{2}\right) =0.  \label{2112b}
\end{eqnarray}
For the case of affine $\widehat{D}_{5}$, we have
\begin{eqnarray}
\widehat{D}_{5} &:&0=y^{2}+x^{3}+z^{6}+xyz  \notag  \label{2113a} \\
&&+v\left( b_{1}yz^{5}+b_{2}z^{8}+c_{1}yxz^{3}+c_{2}x^{2}z^{4}\right)
\label{2113} \\
&&+v^{2}\left( a_{1}z^{10}+a_{2}xz^{8}\right)  \notag  \label{2113b}
\end{eqnarray}
The\ generic relations for eq(\ref{210}) with generic finite and affine $%
\widehat{ADE}$ singularities, including eqs(\ref{2112a}), eqs(\ref{2112b}),
eqs(\ref{2113}), may be found in \cite{10,11}, \cite{12}.\ Before going
ahead note that for affine singularities, the powers $l$ of the complex
variable $v$ appearing in the mirror geometries are given by the Dynkin
weights of the corresponding affine algebra. These integers have a
remarkable interpretation in toric geometry representation of CY threefolds.
The vertices $V_{\alpha }=\left( l_{\alpha },r_{\alpha },s_{\alpha }\right) $
are arranged into subsets belonging to parallel square lattices of $Z^{3}$.
For the case of eq(\ref{2113}),\ the vertices of the elliptic curve have $%
l_{\alpha }=0$; those associated with the nodes with Dynkin weights $1$\
have $l_{\alpha }=1$ and finally those Dynkin nodes with weights $2$\ have $%
l_{\alpha }=2$. Note moreover, that homogeneity of the polynomials $P\left(
X^{\ast }\right) $ describing mirror geometries with affine singularities
requires $v$ to have, in general, a homogeneous dimensions $6-N$. This
dimension vanishes for $N=6$, a remarkable feature which give the freedom to
have arbitrary powers of $v$ without affecting the homogeneity of $P\left(
X^{\ast }\right) $. The solutions we will derive here below for the case of
hyperbolic singularities use, amongst others, this special property. To
avoid confusion, from now on, we denote the variable $v$ of the case $N=6$
by $t$ and thinking about it as parameterizing the complex plane.

\subsection{Flat Hyperbolic bundles on elliptic K3}

In the case of hyperbolic singularities we are after, the previous elliptic
curve is replaced by the following special elliptic fibered $K3$ surface,
\begin{equation}
y^{2}+x^{3}+t^{-1}z^{6}+xyz=0,  \label{2114}
\end{equation}
where the fourth variable $t$ parameterizes the base of $\mathbf{K}3$. This
complex surface describes the base manifold of the mirror CY threefold with
hyperbolic singularities we are interested in here. As these kind of
unfamiliar manifolds look a little bit unusual, let us give also the
solutions for the vertices these CY threefolds. For the case of eq(\ref{2114}%
), the corresponding four vertices read as
\begin{eqnarray}
yxz\quad &\longleftrightarrow &\quad \left( 0,0,0\right)  \notag \\
y^{2}\quad &\longleftrightarrow &\quad \left( 0,0,-1\right)  \notag \\
x^{3}\quad &\longleftrightarrow &\quad \left( 0,-1,0\right) \\
t^{-1}z^{6}\quad &\longleftrightarrow &\quad \left( -1,2,3\right)  \notag
\end{eqnarray}
To get the algebraic geometry eq
\begin{equation}
P\left( X_{hyp}^{\ast }\right)
=y^{2}+x^{3}+t^{-1}z^{6}+xyz+\sum_{i=1}^{n}a_{i}y_{i},
\end{equation}
describing mirror geometries of CY threefolds with hyperbolic singularities
we solve eqs(\ref{211}) for hyperbolic generalized Cartan matrices of order $%
n$. The $y_{i}$s one gets are given by monomials of type $y_{i}=y^{\alpha
_{i}}x^{\beta _{i}}z^{\gamma _{i}}t^{l_{i}}$ and so
\begin{equation}
P\left( X_{hyp}^{\ast }\right)
=y^{2}+x^{3}+t^{-1}z^{6}+xyz+\sum_{i=1}^{n}a_{i}\left( w\right) y^{\alpha
_{i}}x^{\beta _{i}}z^{\gamma _{i}}t^{l_{i}}  \label{2116}
\end{equation}
The vertices $V_{i}$ corresponding to the monomials $y^{\alpha _{i}}x^{\beta
_{i}}z^{\gamma _{i}}t^{l_{i}}$ have first entries equal to the $l_{i}$
powers of the variable $t$; i.e,
\begin{equation}
y^{\alpha _{i}}x^{\beta _{i}}z^{\gamma _{i}}t^{l_{i}}\quad
\longleftrightarrow \quad \left( l_{i},r_{\alpha },s_{\alpha }\right) .
\label{2117}
\end{equation}
The remaining others are determined by solving eq $\sum_{\alpha
=0}^{n+4}q_{\alpha }^{i}V_{\alpha }=0$. To see how the machinery works in
practice, let us illustrate our solutions on the four hyperbolic Lie algebra
examples of figure 4.

\subsection{Explicit solutions}

\qquad Here we consider the mirror geometry of CY threefolds with $\mathcal{H%
}_{3}^{4}$, $\mathcal{H}_{1}^{5}$, $\mathcal{H}_{1}^{7}$ and $\mathcal{H}%
_{1}^{8}$ hyperbolic singularities of figures 4. First we describe the
surfaces with $\mathcal{H}_{3}^{4}$ and $\mathcal{H}_{1}^{5}$\
singularities; then consider the case of Calabi-Yau threefolds based on $%
\mathcal{H}_{3}^{4}$ and $\mathcal{H}_{1}^{5}$\ geometries. After, we give
the mirror geometries with gauge groups and fundamental matters that lead,
in the weak gauge coupling regime, to $\mathcal{N}=2$ supersymmetric CFT$%
_{4} $s with $\mathcal{H}_{3}^{4}$ and $\mathcal{H}_{1}^{5}$ singularities.
As the analysis for $\mathcal{H}_{1}^{5}$\ and $\mathcal{H}_{1}^{8}$\ is
quite similar to $\mathcal{H}_{3}^{4}$ and $\mathcal{H}_{1}^{7}$
respectively, we give just the results for these algebras.

\begin{itemize}
\item  \textit{Mirror geometry of surfaces with} $\mathcal{H}_{3}^{4}$
\textit{singularity}
\end{itemize}

To get the mirror geometry of surface with $\mathcal{H}_{3}^{4}$
singularity, we have to solve eqs(\ref{211}). To that purpose, we have to
specify the $q_{\alpha }^{i}$ integers associated with $\mathcal{H}_{3}^{4}$%
. They are given by the following $4\times \left( 4+4\right) $\ rectangular\
matrix

\begin{equation}
q_{\alpha }^{i}=\left(
\begin{array}{cccccccc}
-2 & 1 & 0 & 0 & 0 & 0 & 0 & 1 \\
1 & -2 & 1 & 1 & 0 & 0 & 0 & -1 \\
0 & 1 & -2 & 1 & 3 & 1 & 0 & -4 \\
0 & 1 & 1 & -2 & 0 & 1 & 0 & -1
\end{array}
\right) .  \label{2118}
\end{equation}
The first $4\times 4$ block is just the Cartan matrix for $\mathcal{H}%
_{3}^{4}$\ and the other $3\times 4$ block is associated with the $y_{5}$, $%
y_{6}$, $y_{7}$ and $y_{8}$ realized as,
\begin{equation}
y_{5}=y^{2},\quad y_{6}=x^{3},\quad y_{7}=t^{-1}z^{6},\quad y_{8}=xyz.
\label{2119}
\end{equation}
Putting eqs(\ref{2119}) and (\ref{2118}) back into eqs(\ref{211}), we get,
after some straightforward computations, the following result:
\begin{eqnarray}
P\left( X_{\mathcal{H}_{3}^{4}}^{\ast }\right) &=&y^{2}+x^{3}+z^{6}t^{-1}+xyz
\notag \\
&&+\left[ \mathrm{a}z^{6}+\mathrm{b}tz^{6}+\mathrm{c}txz^{4}+\mathrm{d}%
yz^{3}t\right] ,  \label{2120}
\end{eqnarray}
where $\mathrm{a}$, $\mathrm{b}$, $\mathrm{c}$ and $\mathrm{d}$ are complex
structures. This is a compact homogeneous complex two dimension surface
embedded in $\mathbf{WP}^{3}\left( 3,2,1,0\right) $; it shares features with
affine $A_{2}$ mirror geometry (\ref{2112a}). The vertices associated with
the monomials of the second line of eq(\ref{2110}) are,
\begin{eqnarray}
z^{6}\quad &\longleftrightarrow &\quad \left( 0,2,3\right)  \notag \\
tz^{6}\quad &\longleftrightarrow &\quad \left( 1,2,3\right)  \label{021} \\
txz^{4}\quad &\longleftrightarrow &\quad \left( 1,1,2\right)  \notag \\
tyz^{3}\quad &\longleftrightarrow &\quad \left( 1,1,1\right) .  \notag
\end{eqnarray}
In the case of $\mathcal{H}_{1}^{5}$, the Mori vectors are given by
\begin{equation}
q_{\alpha }^{i}=\left(
\begin{array}{ccccccccc}
-2 & 1 & 0 & 0 & 0 & 0 & 0 & 1 & 0 \\
1 & -2 & 1 & 0 & 1 & 0 & 0 & 1 & -1 \\
0 & 1 & -2 & 1 & 0 & 0 & 1 & 0 & 0 \\
0 & 1 & 0 & -2 & 1 & 2 & 0 & 0 & -2 \\
0 & 0 & 1 & 1 & -2 & 1 & 2 & 0 & -3
\end{array}
\right)
\end{equation}
and eq(\ref{2120}) extends as,
\begin{eqnarray}
P\left( X_{\mathcal{H}_{1}^{5}}^{\ast }\right) &=&y^{2}+x^{3}+z^{6}t^{-1}+xyz
\notag \\
&&+\left[ \mathrm{a}z^{6}+\mathrm{b}tz^{6}+\mathrm{c}txz^{4}+\mathrm{d}%
yz^{3}t+\mathrm{e}tx^{2}z^{2}\right] .
\end{eqnarray}
Here also one can write down the analogue of the vertices (\ref{021}).

\begin{itemize}
\item  \textit{Calabi-Yau threefolds with} $\mathcal{H}_{3}^{4}$ and $%
\mathcal{H}_{1}^{5}$\ \textit{singularities}.
\end{itemize}

To get the mirror of Calabi-Yau threefolds with $\mathcal{H}_{3}^{4}$ and $%
\mathcal{H}_{1}^{5}$\ \textit{singularities}, we have to vary the complex
structures \textrm{f} ( \textrm{f} stands for $\mathrm{a}$, $\mathrm{b}$, $%
\mathrm{c}$, $\mathrm{d}$ and $\mathrm{e}$ ) on the $\mathbf{CP}^{1}$ base
parameterized by $w$%
\begin{eqnarray}
P\left( X_{\mathcal{H}_{3}^{4}}^{\ast }\right) &=&y^{2}+x^{3}+z^{6}t^{-1}+xyz
\notag \\
&&+\left[ z^{6}\mathrm{a}\left( w\right) +tz^{6}\mathrm{b}\left( w\right)
+txz^{2}\mathrm{c}\left( w\right) +yzt\mathrm{d}\left( w\right) \right] ,
\end{eqnarray}
where \textrm{f}$\left( w\right) $s are as in (\ref{8}). A similar eq is
also valid for $\mathcal{H}_{1}^{5}$ hyperbolic singularity. In the $%
\mathcal{N}=2$ supersymmetric QFT$_{4}$ limit, the degree $n_{\mathrm{f}}$
of these polynomials defines the rank of the gauge group $SU\left( n_{%
\mathrm{f}}+1\right) $ fibered on the corresponding node.

\begin{itemize}
\item  \textit{Hyperbolic CFT}$_{4}$\textit{s}
\end{itemize}

The $\mathcal{N}=2$ supersymmetric QFT$_{4}$ limit may be made scale
invariant by introducing fundamental matters. Since, the $\mathbf{K}%
_{ij}n_{j}$ vector should be equal to $-m_{i}$; with $m_{i}$\ positive
integers, we have
\begin{eqnarray}
2n_{a}-n_{b} &=&m_{a}  \notag \\
-n_{a}+2n_{b}-n_{c}-n_{d} &=&m_{b}  \notag \\
-n_{b}+2n_{c}-n_{d} &=&m_{c} \\
-n_{b}-n_{c}+2n_{d} &=&m_{d}.  \notag
\end{eqnarray}
In geometric engineering method, this is equivalent to engineer $m_{i}%
\mathbf{n}_{f_{i}}$ fundamental matters of $SU\left( n_{f_{i}}\right) $ on
the trivalent nodes of the hyperbolic diagram with negative Mori charges.
The simplest configuration one may write down corresponds to take $n_{a}=0$,
$n_{b}=n_{c}=n_{d}=n$ ; i.e ( $m_{a}=n$, $m_{b}=m_{c}=m_{d}=0$\ ) and
engineer a $SU\left( n\right) $\ flavor symmetry on the hyperbolic node by
using trivalent geometry techniques of \cite{10}. The resulting mirror
geometry reads as
\begin{eqnarray}
P\left( X_{\mathcal{H}_{3}^{4}}^{\ast }\right) &=&y^{2}+x^{3}+z^{6}t^{-1}+xyz
\notag \\
&&+\left[ \frac{1}{\mathrm{\alpha }\left( w\right) }z^{6}+tz^{6}\mathrm{b}%
\left( w\right) +txz^{4}\mathrm{c}\left( w\right) +yzt\mathrm{d}\left(
w\right) \right]
\end{eqnarray}
where $\mathrm{\alpha }\left( w\right) $, $\mathrm{b}\left( w\right) $, $%
\mathrm{c}\left( w\right) $ and $\mathrm{d}\left( w\right) $ are all of them
polynomials of degree $n$. A quite similar result is valid for $\mathcal{N}%
=2 $ CFT$_{4}$ based on the $\mathcal{H}_{1}^{5}$\ singularity.

\textbf{CY threefolds with}\textit{\ }$\mathcal{H}_{1}^{7}$\textit{\ }%
\textbf{and}\textit{\ }$\mathcal{H}_{1}^{8}$\textit{\ }\textbf{singularities}

These solutions share features with affine $\mathcal{D}_{5}$ and affine $%
\mathcal{D}_{6}$ mirror geometries of \cite{10}. We will derive here the
mirror geometry associated with $\mathcal{H}_{1}^{7}$\ singularity and the
underlying $\mathcal{N}=2$ supersymmetric CFT$_{4}$. Then we give the
results for $\mathcal{H}_{1}^{8}$.

\begin{itemize}
\item  \textit{Surfaces with} $\mathcal{H}_{1}^{7}$ and $\mathcal{H}_{1}^{8}
$ singularities
\end{itemize}

Complex surface with $\mathcal{H}_{1}^{7}$ singularity is described by
\begin{equation}
P\left( X_{hyp}^{\ast }\right)
=y^{2}+x^{3}+t^{-1}z^{6}+xyz+\sum_{i=1}^{7}a_{i}y_{i}=0,  \label{h17}
\end{equation}
where the the $y_{i}$s are constrained as $\prod_{j=1}^{7}y_{j}^{q_{j}^{i}}=%
\prod_{\alpha =8}^{11}y_{\alpha }^{-q_{\alpha }^{i}}$ and where $q_{j}^{i}$
is given by,
\begin{equation}
q_{\alpha }^{i}=\left(
\begin{array}{ccccccccccc}
-2 & 1 & 0 & 0 & 0 & 0 & 0 & 0 & 0 & 1 & 0 \\
1 & -2 & 1 & 0 & 0 & 0 & 0 & 0 & 0 & 0 & 0 \\
0 & 1 & -2 & 1 & 1 & 0 & 0 & 0 & 0 & 0 & -1 \\
0 & 0 & 1 & -2 & 0 & 0 & 0 & 1 & 0 & 0 & 0 \\
0 & 0 & 1 & 0 & -2 & 1 & 1 & 0 & 0 & 0 & -1 \\
0 & 0 & 0 & 0 & 1 & -2 & 0 & 2 & 1 & 0 & -2 \\
0 & 0 & 0 & 0 & 1 & 0 & -2 & 0 & 1 & 0 & 0
\end{array}
\right) .  \label{h71}
\end{equation}
Using this explicit expression (\ref{h71}), one can solve the expression of
the $y_{i}$s in terms of $x,y,z$ and $t$. Putting these expressions back
into (\ref{h17}), we get the following result:
\begin{eqnarray}
P\left( X_{\mathcal{H}_{1}^{7}}^{\ast }\right) &=&y^{2}+x^{3}+z^{6}t^{-1}+xyz
\notag \\
&&+\mathrm{h}z^{6}+\mathrm{b}_{2}tz^{6}+\mathrm{b}_{2}yz^{3}t+\mathrm{c}%
_{1}txyz+\mathrm{c}_{2}txz^{2} \\
&&+\mathrm{a}_{1}t^{2}z^{6}+\mathrm{a}_{2}t^{2}xz^{4},  \notag
\end{eqnarray}
where $\mathrm{a}_{i}$, $\mathrm{b}_{i}$, $\mathrm{c}_{i}$ and $\mathrm{h}$
are complex moduli. For the case of $\mathcal{H}_{1}^{8}$\ singularity, the
result we get is
\begin{eqnarray}
P\left( X_{\mathcal{H}_{1}^{8}}^{\ast }\right) &=&y^{2}+x^{3}+z^{6}t^{-1}+xyz
\notag \\
&&+\mathrm{h}z^{6}+\mathrm{b}_{2}tz^{6}+\mathrm{b}_{2}yz^{3}t+\mathrm{c}%
_{1}txyz+\mathrm{c}_{2}tx^{3} \\
&&+\mathrm{a}_{1}t^{2}z^{6}+\mathrm{a}_{2}t^{2}xz^{4}+\mathrm{a}%
_{3}t^{2}x^{2}z^{2}.  \notag
\end{eqnarray}

\begin{itemize}
\item  \textit{Calabi-Yau threefolds with} $\mathcal{H}_{1}^{7}$ and $%
\mathcal{H}_{1}^{8}$\ \textit{singularities}
\end{itemize}

Varying the complex moduli on $\mathbf{CP}^{1}$ base, we get the mirror of
CY threefolds with $\mathcal{H}_{1}^{7}$ and $\mathcal{H}_{1}^{8}$\ \textit{%
singularities}. We obtain
\begin{eqnarray}
P\left( X_{\mathcal{H}_{1}^{7}}^{\ast }\right) &=&y^{2}+x^{3}+z^{6}t^{-1}+xyz
\notag \\
&&+\mathrm{h}\left( w\right) z^{6}+\mathrm{b}_{2}\left( w\right) tz^{6}+%
\mathrm{b}_{1}\left( w\right) yz^{3}t+\mathrm{c}_{1}\left( w\right) txyz+%
\mathrm{c}_{2}\left( w\right) tx^{3} \\
&&+\mathrm{a}_{1}\left( w\right) t^{2}z^{6}+\mathrm{a}_{2}\left( w\right)
t^{2}xz^{4}+\mathrm{a}_{3}\left( w\right) t^{2}x^{2}z^{2},  \notag
\end{eqnarray}
and a quite similar eq for $\mathcal{H}_{1}^{8}$.

\begin{itemize}
\item  \textit{Hyperbolic CFT}$_{4}$\textit{s}
\end{itemize}

The $\mathcal{N}=2$ supersymmetric QFT$_{4}$ limit may be made scale
invariant by introducing fundamental matters. As before the $\mathbf{K}%
_{ij}n_{j}$ vector,
\begin{equation}
\left(
\begin{array}{ccccccc}
2 & -1 & 0 & 0 & 0 & 0 & 0 \\
-1 & 2 & -1 & 0 & 0 & 0 & 0 \\
0 & -1 & 2 & -1 & -1 & 0 & 0 \\
0 & 0 & -1 & 2 & 0 & 0 & 0 \\
0 & 0 & -1 & 0 & 2 & -1 & -1 \\
0 & 0 & 0 & 0 & -1 & 2 & 0 \\
0 & 0 & 0 & 0 & -1 & 0 & 2
\end{array}
\right) \left(
\begin{array}{c}
n_{h} \\
n_{b_{2}} \\
n_{a_{1}} \\
n_{b_{1}} \\
n_{a_{2}} \\
n_{c_{1}} \\
n_{c_{2}}
\end{array}
\right) =\left(
\begin{array}{c}
2n_{h}-n_{b_{2}} \\
-n_{h}+2n_{b_{2}}-n_{a_{1}} \\
-n_{b_{2}}+2n_{a_{1}}-n_{b_{1}}-n_{a_{2}} \\
-n_{a_{1}}+2n_{b_{1}} \\
-n_{a_{1}}+2n_{a_{2}}-n_{c_{1}}-n_{c_{2}} \\
-n_{a_{2}}+2n_{c_{1}} \\
-n_{a_{2}}+2n_{c_{2}}
\end{array}
\right) ,
\end{equation}
should be a negative integer vector; i.e $\mathbf{v}=-\mathbf{m}$. Its
entries are interpreted as the numbers of fundamental matters one should
engineer on the trivalent nodes of $\mathcal{H}_{1}^{7}$\ with negative Mori
vector weight. There are various kinds of solutions one can write down; but
the simplest one corresponds to take

\begin{eqnarray}
n_{h} &=&0,  \notag \\
n_{b_{1}} &=&n_{b_{1}}=n_{c_{1}}=n_{c_{2}}=n, \\
n_{a_{1}} &=&n_{a_{2}}=2n,  \notag
\end{eqnarray}
which correspond to have $m_{h}=n$\ and $%
m_{b_{1}}=m_{b_{1}}=m_{c_{1}}=m_{c_{2}}=m_{a_{1}}=m_{a_{2}}=0$. Therefore,
the full symmetry of this $\mathcal{N}=2$ hyperbolic CFT$_{4}$ has $%
SU^{4}\left( n\right) \times SU^{2}\left( 2n\right) $ as a gauge group and $%
SU\left( n\right) $ flavor symmetry carried by the negative node of the
trivalent vertex of the $\mathcal{H}_{1}^{7}$ hyperbolic chain. The
corresponding mirror geometry is described by the following algebraic eq
\begin{eqnarray}
P\left( X_{\mathcal{H}_{1}^{7}}^{\ast }\right) &=&y^{2}+x^{3}+z^{6}t^{-1}+xyz
\notag \\
&&+\frac{1}{\mathrm{\alpha }\left( w\right) }z^{6}+\mathrm{b}_{2}\left(
w\right) tz^{6}+\mathrm{b}_{1}\left( w\right) yz^{3}t+\mathrm{c}_{1}\left(
w\right) txyz+\mathrm{c}_{2}\left( w\right) tx^{3} \\
&&+\mathrm{a}_{1}\left( w\right) t^{2}z^{6}+\mathrm{a}_{2}\left( w\right)
t^{2}xz^{4}+\mathrm{a}_{3}\left( w\right) t^{2}x^{2}z^{2}  \notag
\end{eqnarray}
A similar result may be also written down for $\mathcal{N}=2$ CFT$_{4}$
based on the $\mathcal{H}_{1}^{8}$\ singularity.

\section{Conclusion and Outlook}

\qquad In this paper, we have derived a new class of $\mathcal{N}=2$ CFT$%
_{4} $s embedded in type IIA superstring on CY threefolds with indefinite
singularities. We have also given the geometric engineering of $\mathcal{N}%
=2 $ QFT$_{4}$ based on a special set of hyperbolic singularities as well as
their infrared limit. This development has consequences on the understanding
$\mathcal{N}=2$ QFT$_{4}$s embedded in type IIA string theories; but also on
the classification of indefinite singularities of complex surfaces. For the
first point and besides the fact that most of the results, obtained in the
context of ADE geometries during the last few years, extends naturally to
the indefinite hyperbolic sector, there is also a remarkable and new fact
which appear in this kind of field theories. Indeed viewed from type II
superstring compactifications, our result tells that one may also have non
perturbative gauge symmetries engineered by Lie algebras of indefinite type.
It would be an interesting task to deeper this issue. From algebraic
geometry point of view, our present study gives the first explicit steps
towards the understanding of Calabi-Yau manifolds with indefinite
singularities. Though type IIA defining algebraic geometry eqs of such
singularities are still lacking, one may usually get precious informations
on them by using the mirror geometry eqs
\begin{equation}
\prod_{j=1}^{n}y_{j}^{K_{ij}}=\prod_{\alpha \geq 1}\xi _{\alpha
}^{-q_{\alpha }^{i}}.
\end{equation}
where $\mathbf{K}_{ij}$ is the generalized Cartan matrix. \ From these eqs,
one may solve the $y_{i}$s in term of $\xi _{\alpha }$; but this program
requires what $\xi _{\alpha }$s are? In the case of \textit{finite} $ADE$
singularities, these $\xi _{\alpha }$s parameterize a real two sphere $%
S^{2}\sim \mathbf{CP}^{1}$ and for \textit{affine} $ADE$; they parameterize
an elliptic curve embedded in $\mathbf{WP}^{2}$. For the \textit{indefinite}
( hyperbolic ) singularities, we have learnt from our present work that $\xi
_{\alpha }$s\ \ parameterize the ( elliptic ) $K3$ surface $%
y^{2}+x^{3}+z^{6}t^{-1}+xyz$ with the remarquable pole at $t=0$. We suspect
that this is a signature for\ complex surfaces with indefinite
singularities. We also suspect that, instead of the simple poles encountered
the examples we have considered in this paper, one may have in general
higher orders poles as well. More details regarding this special issue are
considered in \cite{23}.

\qquad We end this conclusion by discussing a link between the Lie algebraic
solutions we have developed in the present study and representation theory
of the gauge quiver diagrams. A naive way to see how representations of
quiver graphs can be implemented in our analysis is to think about the three
relations,
\begin{equation}
\sum_{j=1}^{r}\mathbf{K}_{ij}^{\left( q\right) }n_{j}=qm_{i};\qquad
q=0,+1,-1,  \label{c1}
\end{equation}
as the Lie algebraic set up behind some representation theory identities
carrying same informations. Results from KM algebra representations show
that the identities in question are given by the duality property between
positive simple roots $a_{\nu }$ and fundamental weights $\Lambda _{\nu }$
which, we prefer to write it, for generalised symmetrisable Cartan matices $%
\mathbf{K}_{\mu \nu }$ with $corank\left( \mathbf{K}_{\mu \nu }\right) =0$
and order $\left( r+l+1\right) $, as follows,
\begin{equation}
\sum_{\nu =-l}^{r}\mathbf{K}_{\mu \nu }\Lambda _{\nu }=a_{\nu };\qquad \mu
=-l,...,-1,0,1,2,...,r.  \label{c2}
\end{equation}
Notice that the duality identity between $\Lambda _{\nu }$ and $a_{\nu }$
reads in general $<\Lambda _{\mu },a_{\nu }>=\delta _{\mu \nu }$ with $<,>$
is the non degenerate symmetric bilinear form on KM algebra. For invertible $%
\mathbf{K}_{\mu \nu }$s, this duality eq can be usually put in the above
form. To get the constraint relation\ describing the passage between eqs (%
\ref{c1}) and eqs(\ref{c2}), we will restrict ourselves here to give the
main lines of the method which we illustrate below on the rank $\left(
r+2\right) $ hyperbolic Lie algebras subset we have been studying.

\qquad To that purpose, recall first that the commutation relations of
hyperbolic algebras $\mathcal{H}^{\left( r+2\right) }$, with Chevalley
generators $e_{\mu },$ $f_{\mu },$ $h_{\mu }$; $-1\leq \mu \leq r$, are
\begin{eqnarray}
\left[ \mathrm{h}_{\mu },\mathrm{e}_{\mu }\right]  &=&\mathcal{K}_{\mu \nu }%
\mathrm{e}_{\nu };\qquad \left[ \mathrm{h}_{\mu },\mathrm{f}_{\nu }\right] =-%
\mathcal{K}_{\mu \nu }\mathrm{f}_{\nu };\qquad \left[ \mathrm{e}_{\mu },%
\mathrm{f}_{\nu }\right] =\delta _{\mu \nu }\mathrm{h}_{\mu };  \notag \\
\left( ad\mathrm{e}_{\mu }\right) ^{1-\mathcal{K}_{\mu \nu }}\mathrm{e}_{\nu
} &=&\left( ad\mathrm{f}_{\mu }\right) ^{1-\mathcal{K}_{\mu \nu }}\mathrm{f}%
_{\nu }=0;\qquad ;\mu \neq \nu ,\qquad -1\leq \mu ,\nu \leq r  \label{c3}
\end{eqnarray}
where $h_{i};1\leq i\leq r$ are the usual commuting Cartan generators and
the two extra $h_{0}$ and $h_{-1}$ are as: $h_{0}=k-\sum_{i=1}^{r}\widehat{%
\mathrm{s}}_{i}h_{i}$ and $h_{-1}=-k-d$ with $k$ and $d$ respectively the
central element and the derivation of the affine subalgebra $\widehat{%
\mathrm{g}}_{r}$ of $\mathcal{H}^{\left( r+2\right) }$ \cite{24}-\cite{27}.
The $\widehat{\mathrm{s}}_{i}$s are roughly speaking the Dynkin weights and $%
\mathrm{k}$ and $d$ generators have scalar products given by $<k\mathrm{,}%
k>=<d,d>=0$ and $<k,d>=1$. Second, note that like in ordinary and affine Lie
algebras, the $\left( r+2\right) $ fundamental weights $\Lambda _{\mu }$ and
positive simple roots $a_{\mu }$ of hyperbolic algebras are immediately
deduced by extending the $\left( r+1\right) $ affine ones. In addition to
the ordinary positive simple roots $a_{i}=\alpha _{i}$; $i=1,...,r$, there
are two extra ones namely $a_{-1}$ and $a_{0}$ which read as $a_{-1}=-%
\mathrm{k}-\delta $, $a_{0}=\delta -\psi $ with $\psi
=\sum_{i=1}^{r}s_{i}\alpha _{i}$\ being the usual highest root. Similarly,
we have for the fundamental weights the realisation $\Lambda _{-1}=-\delta $%
, $\Lambda _{0}=\mathrm{k}-\delta $ and $\Lambda _{i}=\lambda _{i}+s_{i}%
\mathrm{k}-s_{i}\delta $. Both of these $a_{\mu }$ and $\Lambda _{\mu }$
satisfy the duality property and may be rewritten in a more convenient form
as extended objects like,
\begin{eqnarray}
a_{-1} &=&\left( 0,-1,-1\right) ;\qquad a_{0}=\left( -\psi ,0,1\right)
;\qquad a_{i}=\left( \alpha _{i},0,0\right) ,\quad i=1,...,r,  \notag \\
\Lambda _{-1} &=&\left( 0,0,-1\right) ;\qquad \Lambda _{0}=\left(
0,1,-1\right) ;\qquad \Lambda _{i}=\left( \lambda _{i},s_{i},-s_{i}\right)
,\quad i=1,...,r,  \label{c4}
\end{eqnarray}
where the two extra directions refer to the $\mathrm{k}$ and $\delta $
generators. To bring eq(\ref{c2}) into eq(\ref{c1}), we use a nice property
of the Lorentzian structure of the root lattice $\mathcal{L}_{r+2}$ for
hyperbolic $\mathcal{H}^{\left( r+2\right) }$ and do it in three steps by
performing projections of eq(\ref{c2}) along root lattice vectors $\mathbf{r}%
_{q}$ of the three regions of $\mathcal{L}_{r+2}$ namely \textit{space} like
vectors ($\mathbf{r}_{+}^{2}>0$), \textit{light} like vectors ($\mathbf{r}%
_{0}^{2}=0$) and \textit{time} like vectors ($\mathbf{r}_{-}^{2}<0$). These
three sectors turns out to correspond exactly to the three relations of eqs(%
\ref{c1}) we are after. We first treat the case of finite ADE algebras, then
affine ADE and finally their hyperbolic over extension. (\textbf{a}) For
finite ADE, the representation identity involving ordinary $\alpha _{i}$s
and $\lambda _{i}$s and contained\footnote{%
Eqs(\ref{c5}) for ADE subalgeras of $\mathcal{H}^{\left( r+2\right) }$ can
be rederived under projection of eqs(\ref{c2}) along r$_{+}$.} in eq(\ref{c2}%
), reads as,
\begin{equation}
\sum_{j=1}^{r}\mathbf{K}_{ij}^{\left( +\right) }\lambda _{j}=\alpha
_{i};\qquad i,j=1,2,...,r,  \label{c5}
\end{equation}
where $\mathbf{K}_{ij}^{\left( +\right) }$ is as before. To put this
relation into the form $\mathbf{K}_{ij}^{\left( +\right) }n_{j}=m_{i}$, we
consider a positive integer space like vector $\mathbf{r}_{+}\mathbf{=}%
\sum_{i=1}^{r}n_{i}\alpha _{i}$ of the ADE\ root lattice such that $<\mathbf{%
r}_{+},\alpha _{i}>=m_{i}$ is a positive integer but $<\mathbf{r}%
_{+},\Lambda _{\mu }>=<\mathbf{r}_{+},\alpha _{\mu }>=0$ for $\mu =-1,0$.
Projecting eq(\ref{c5})\ along this vector $\mathbf{r}_{+}$, we get the
desired eq(\ref{1}); thanks to the duality relation $<\alpha _{i},\lambda
_{j}>=\delta _{ij}$. Therefore, we learn that the ranks $n_{i}$ of the
quiver gauge group $\prod_{i=1}^{r}U\left( n_{i}\right) $ are just the
projections of $\mathbf{r}_{+}$ on the $\lambda _{i}$ fundamental weights
and the numbers $m_{i}$ of fundamental matters are the projections of $%
\mathbf{r}_{+}$ on the $\alpha _{i}$ simple roots. (\textbf{b}) For affine
case, we get a similar interpretation in term of representation theory by
taking the projection of $\mathbf{K}_{\mu \nu }\Lambda _{\nu }=a_{\nu }$
along the imaginary root lattice vector $\mathbf{r}_{0}=n\delta $. Using the
identities $<\mathbf{r}_{0},\alpha _{\mu }>=-n\delta _{-1,\mu }$ and $<%
\mathbf{r}_{0},\Lambda _{-1}>=0$, $<\mathbf{r}_{0},\Lambda _{0}>=n$ and $<%
\mathbf{r}_{0},\Lambda _{i}>=ns_{i}$ as one may check from eqs(\ref{c4}),
one obtains two eqs, one of which is trivial and the second is just $\mathbf{%
K}_{ij}^{\left( 0\right) }n_{j}=0$ eq(\ref{2}) with the right integer values
$n_{j}=ns_{j}$. Note that the absence of fundamental matter in affine case
has a nice geometric interpretation in hyperbolic root lattice; it is
manifested in the present approach by the remarkable property $<\mathbf{r}%
_{0},a_{i}>=<\mathbf{r}_{0},\delta >=0$. (\textbf{c}) Finally considering a
positive integer time like vector $\mathbf{r}_{-}=\sum_{i=-1}^{r}n_{i}a_{i}$
of the hyperbolic root lattice having projections along simple roots given
by negative integers; i.e $<\mathbf{r}_{-},\alpha _{i}>=-m_{i}$, and
projecting eq(\ref{c2}) along $\mathbf{r}_{-}$, we get the desired result
once more. Here also the $n_{i}$s are the projections of $\mathbf{r}_{-}$ on
the fundamental weights and the number of fundamental matter is the opposite
of its projections on simple roots. More details on this representation
analysis as well as extensions and applications in supersymmetric quiver
gauge theories will be considered in a future occasion.

\begin{acknowledgement}
This work is supported by Protars III, CNRST, Rabat, Morocco. A.Belhaj is
supported by Ministerio de Education cultura y Deporte, grant SB 2002-0036.
\end{acknowledgement}

\end{document}